\documentclass[conference]{IEEEtran}
\usepackage{caption}
\usepackage{cite}
\usepackage{graphics}
\usepackage{graphicx}
\usepackage{amsmath,amsfonts,amssymb}
\usepackage{blindtext, graphicx}
\usepackage{mathtools}
\usepackage{subcaption}
\usepackage[ruled,vlined,lined,boxed,commentsnumbered]{algorithm2e}
\begin{document}

\title{Meeting Energy-Efficient and QoS Requirements of 5G Using D2D Communications}

\author{
\IEEEauthorblockN{Jean-Marc Kelif}
\IEEEauthorblockA{Orange Labs\\
France\\
Email: jeanmarc.kelif@orange.com}
\and
\IEEEauthorblockN{William Diego}
\IEEEauthorblockA{Orange Labs\\
France\\
Email: william.diego@orange.com}
}

\maketitle

\begin{abstract}

Device-to-device (D2D) communication is a promising technology for the future wireless systems. It allows direct communication between devices, which provides improvements in terms of delay, throughput and energy consumption. Therefore, it can contribute to achieving  the ambitious requirements of future 5G wireless system. In this sense, energy efficiency has become a key requirement in the design of 5G technology. In this paper we analyze the energy-efficiency improvement provided by D2D communications in an overlaying scenario, in the context of a realistic wireless network system. This analysis takes into account the two D2D phases, discovery and communication.  A centralized architecture is considered to manage discovery, which is a key phase on D2D communications. Numerical evaluation shows improvement in terms of energy-efficiency, reachable throughput and outage probability.   


\end{abstract}

\begin{IEEEkeywords}
D2D, Energy, Energy-efficiency,  5G, QoS
\end{IEEEkeywords}
\newcommand*{\bufferspace}{1.5mm}
\IEEEpeerreviewmaketitle

\section{Introduction}

Device-to-device (D2D) communication is the technology allowing direct communication between nearby mobile devices, which was introduced in LTE in 3GPP Release 12~\cite{3gppd2d}. D2D communication is recognized as one of the technology components of the evolving 5G architecture. Its potential gains include: reduction of devices transmission power, reduction of communication delay, improvement of spectral efficiency and extension of cell coverage area. D2D communications will guarantee the ubiquity of high Quality of Service (QoS) and offload the infrastructure of mobile network.\\\indent In this context, D2D will become a key technology to meet the 5G's requirements~\cite{alliance20155g} in terms of latency ($<$10ms end-to-end in general) and data rate (at least 50Mbit/s everywhere). \\\indent Moreover, one of the key 5G Radio Access Network (RAN) design requirements pointed by 5GPPP METIS project~\cite{metis2016deliverableD.2.2} is that network-controlled D2D should be natively supported. Therefore, devices will play a key role since they will be integrated into the mobile network as a extension of it. Thus, 5G devices will may act both as terminal and as mobile infrastructure node which means that they will be able to provide services and/or applications~\cite{su2015content,garcia2015edge}. \\\indent Furthermore, the energy-efficiency is a critical requirement of 5G wireless system~\cite{alliance20155g}. In this sense D2D communication is a key technology which will allow to meet energy-efficiency requirements.  \\\indent In this work, we study the potential improvement provided by D2D communications overlaying cellular network in terms of energy-efficiency in the context of a centralized D2D architecture. The proposed centralized D2D architecture aims at improving D2D discovery phase. The major contributions of this article is to analyse the gain of D2D communication overlaying cellular network in terms of  energy-efficiency in 5G mobile networks. The discovery is a key phase in the overall D2D communication procedure, since it determines the pairing configuration of devices in the system. The pairing configuration plays a critical role in the interference management, which directly impacts the overall energy-efficiency and QoS performance. In this regard, this paper intends to analyse both discovery and communication phases in order to give a complete vision of D2D communications performance. This analysis is performed in a centralized D2D system which includes the two D2D phases, discovery and communication phases. An opportunistic algorithm is also presented which aims at improving D2D pairing process (discovery phase).

\section{Background and Motivations}

\subsection{Device-to-Device}

Discovery and communication are the two basic phases to perform a D2D communication. Before the radio resource allocation and data exchange between paired devices, a discovery phase has to be performed. \\\indent During the discovery phase, devices which have D2D capabilities need to identify other similar devices in order to evaluate the possibility to establish D2D communications. In this regard, discovery signals (in predefined subcarriers) are exchanged between devices in order to identify the presence of possible devices in proximity. The quality of link between devices are measured thanks to the discovery signals, which is characterized by the Channel State Information (CSI). Existing works related to the D2D discovery phase can be classified into centralized and distributed approaches. \\\indent In a centralized D2D architecture, a mobile network entity (e.g. D2D controller) manages the set up of D2D connection based on measurements provided by devices (i.e. CSI) \cite{Lin2016device,hong2013analysis}. This entity exploits provided CSIs in order to pair devices according to the operator strategy (e.g. QoS requirements). \\\indent Once the discovery phase was performed and the D2D pairs were established, the communication phase starts. In this phase, the spectrum allocated for the regular communications (between serving base station (BS) and Device) is used by the D2D communications as well. Therefore, one option is to overlay D2D communications over the cellular users (D2D overlay), which means that the available cellular spectrum is partitioned in such a way that the D2D communications and the regular communications use non-overlapping portions of the spectrum. A second option is to underlay D2D communications with respect to regular communications (D2D underlay), which means that the interference management becomes a major challenge. \\\indent In this paper we study the energy-efficiency and QoS performance of overlaying D2D communications taking into consideration a centralized D2D discovery entity, which implements an opportunistic algorithm. The analysis takes into consideration both discovery and communication phases. Numerical evaluation taking into consideration realistic scenarios is also presented in order to evaluate the improvements provided by D2D technology and our discovery opportunistic algorithm.





\subsection{Related Work}

There is a large number of recent papers addressing D2D communication, as evidenced by the following surveys~\cite{asadi2014survey,mach2015band,Noura2016}. However, to the best of our knowledge, there are few existing researches investigating the gains in terms of energy efficiency of D2D communications in 5G's mobile network scenarios taking into consideration both discovery and communication phases. \\\indent In \cite{Zhou2014}, the authors investigate the tradeoff between energy efficiency and spectral efficiency in D2D communications underlaying cellular networks with uplink channel reuse. They propose a distributed energy-efficient resource allocation algorithm for D2D communications. \\\indent In \cite{Wang2013}, the authors propose a resource allocation scheme to improve throughput and battery lifetime. They show that D2D communications underlaying cellular networks can greatly extend device battery lifetime compared with traditional cellular communication. \\\indent In \cite{qiu2013}, the authors analyze energy efficiency of D2D  communication underlaying cellular networks. The results demonstrate that it is more energy efficient if  users communicate with each other directly using D2D communication. \\\indent In \cite{jung2012}, the authors propose a power-efficient discovery strategy and power allocation scheme for D2D communication underlaying cellular networks. The proposed solution shows interesting performance in terms of power-efficiency. \\\indent In \cite{Yaacoub2013}, the authors explore the cooperation between D2D and BS sleeping strategy in order to reduce energy consumption. They show that the combined method led to energy savings for both the devices and the mobile network operator with low impact in QoS in terms of throughput and delay. \\\indent References cited only focus their analysis in the communication phase. In contrast to listed works, this paper focuses on performance evaluation of both discovery and communication phases of D2D communications overlaying cellular network, in order to satisfy 5G requirements in terms of QoS and energy efficiency. We take into consideration a 5G wireless network and analyze the two associated procedures involved in D2D communications, discovery phase and communication phase, in the context of a realistic scenario with 3 sectored sites and 3D antennas in our numerical evaluation.

\section{D2D System Model}

\begin{figure}[!ht]
\centering
\includegraphics[width=7.5cm]{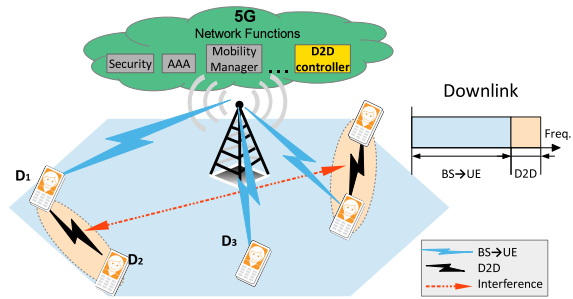}
\caption{\footnotesize Interference scenario for D2D communications overlaying cellular network that use part of frequency bandwidth (DL)}
\vspace{-1.0em}
\label{fig:int}
\end{figure}

We consider a wireless network consisting of $M$ geographical sites, each one composed by 3 BSs. Each antenna covers a sectored cell. We also consider a set of devices which are uniformly and regularly distributed over the two-dimensional plane. D2D is supported natively by all devices. \\\indent We consider that D2D and regular communications are performed based on the Orthogonal Frequency Division Multiple Access (OFDMA) technique. We focus on the downlink (DL), where  D2D communications use dedicated radio resources (i.e. frequency band) as shown in Fig.~\ref{fig:int}.  \\\indent The procedure of pairing (discovery phase) is managed by a centralized entity (i.e. D2D controller). All devices provide signal-to-interference-plus-noise ratio (SINR) of all potential D2D pairs to the D2D controller.  \\\indent We consider that a D2D communication is performed by only two devices. Moreover, once D2D connection is established, one of each D2D pair has to be connected to the BS and acts as a relay node in order to receive data from mobile network as shown in Fig.~\ref{fig:int}. 

The D2D controller is in charge to manage D2D pairing. It explores all possible combinations of D2D pairs and takes a decision based on some defined strategies. In this paper, we present an opportunistic strategy which aims at improving the achievable throughput for each D2D link. We study the scenario where a unique  D2D link between devices is allowed and one of them has to be connected to the BS. \\\indent Once D2D pairs are established, the D2D communication phase can start. Since the dedicated DL bandwidth is simultaneously used by all D2D links, interference management is crucial to achieve good levels of QoS. The interference management is take into consideration by D2D controller during the discovery phase.


\subsection{D2D Discovery Phase (pairing)}

In the proposed D2D discovery phase, each device computes the SINR of each potential pair. The SINR measurement is based on exchanged discovery signals, which are  periodically transmitted on predefined OFDM subcarriers (e.g. Physical Sidelink Discovery Channel - PSDCH~\cite{3gppd2dsig}). Then, a short list of measured SINRs is transmitted to the D2D controller by each device via the control plan. The D2D controller takes the pairing decision based on operator strategy (i.e. SINR over a given threshold). In this paper we implement an opportunistic D2D pairing strategy which aims at improving the achievable throughput for each D2D link, which is described below (Algorithm~\ref{Algo:d2d1}). Finally, the D2D controller triggers signaling procedure to establish D2D connections. \\\indent The D2D controller also selects the device which will act as the relay node (BS$\rightarrow$\underline{\textbf{Device}}$\leftrightarrow$Device). The device in bold means that it is directly connected to the serving BS and plays  the role of relay to the other device. Therefore, each device also transmits to the D2D controller the SINR of its serving BS. This information is used by the D2D controller to select the relay node. \\\indent The mathematical representation of the D2D discovery phase described below is performed by each device in order to calculate SINR related to all potential D2D pairs as well as its serving BS. 

Let us consider:

\begin{itemize}
    \item  $\mathcal{D} = \{1, \ldots , D\}$ the set of devices, uniformly and regularly distributed over the two-dimensional plane.
    \item  $\mathcal{M} = \{1, \ldots , M\}$ the set of BSs, uniformly and regularly distributed over the two-dimensional plane. 
    \item $\mathcal{K} = \{1, \cdots , K\}$ the set of available channels.
    \item $\mathcal{L} = \{1, \ldots , L\}$ the set of D2D links in the network.
    \item $P^{(v)}_{k} (u)$ the transmitted power from the device $v$ to device $u$ over the channel $k$.
    \item $g^{(v)}_{k} (u)$ the propagation channel between transmitter device $v$ and receiver $u$ over the channel $k$.
    \item $P_{bs,k}^{(i)} (u)$ the transmitted power from the BS $i$ to device $u$ over the channel $k$.
    \item $g_{bs,k}^{(i)} (u)$ the propagation channel between BS $i$ and device $u$ over the channel $k$.
\end{itemize}
Since D2D communication use dedicated radio resources (i.e. frequency band - Fig.~\ref{fig:int}), the total amount of power received by a device $u$ from the device $v$ over the discovery channel $k$ is given by the sum of: useful signal $P^{(v)}_{k} (u) g^{(v)}_{k}(u)$, interference due to the other potential D2D links $\sum_{d \in \mathcal{D}, d \neq u,v} P^{(d)}_{k} (u) g^{(d)}_{k}(u)$ and thermal noise $N_{th}$.


Thus, the SINR $\gamma_{k}^{(v)}(u)$ can be expressed as follows:
\begin{equation}
\label{eq:sinr1}
\gamma_{k}^{(v)}(u) = \frac{P^{(v)}_{k}(u) g^{(v)}_{k}(u)}{ \displaystyle\sum_{d \in \mathcal{D}, d \neq u,v} P^{(d)}_{k} (u) g^{(d)}_{k}(u) + N_{th}  }
\end{equation}
The received SINR at device $u$ from a serving BS $i$ can be expressed as follows:
\begin{equation}
\label{eq:sinr2}
\theta_{k}^{(i)}(u) = \frac{P_{bs,k}^{(i)}(u) g_{bs,k}^{(i)}(u)} {\displaystyle\sum_{j\in\mathcal{M},j \neq i}P_{bs,k}^{(j)} (u) g_{bs,k}^{(j)}(u) + N_{th}}
\end{equation}
Then, the SINR of the serving BS for a device $u$ transmitting over the discovery channel $k$ can be expressed as $\theta_{k}^{\prime}(u)$. \\\indent Once the D2D controller has received measurements, the Algorithm~\ref{Algo:d2d1} is used for peering devices. In order to allocate a channel $k$ to a D2D link, the required SINR threshold $\delta_{\text{D2D}}$ needs to be satisfied by the two devices. The threshold $\delta_{\text{D2D}}$ allows to guarantee a minimum throughput level. \\\indent In Algorithm~\ref{Algo:d2d1}, a device $u$ sends to the D2D controller the  SINR of its potential D2D pairs $\gamma_{k}^{(v)}(u)\ \forall\ v \in \mathcal{D}$. And it also sends the SINR of its serving server $\theta_{k}^{\prime}(u)$. Then, the D2D controller evaluates each potential D2D connection $j \in \mathcal{D}$. First, it explores each measured SINR between devices $u$ and $v$ ($u\leftarrow v$ and $u  \rightarrow v$) and chooses the minimal SINR ($\gamma_{\text{D2D}}$), in order to compare it to the required SINR threshold $\delta_{\text{D2D}}$. If $\gamma_{\text{D2D}}$ is lower than $\delta_{\text{D2D}}$, a D2D connection is not possible, otherwise a D2D connection is established. After that, the device having the best connection to its serving BS ($\theta_{\text{BS}}$) is selected as the relay node. This evaluation takes into consideration the SINR measurement $\theta_{k}^{\prime}(u)$ and $\theta_{k}^{\prime}(v)$. Finally, the D2D controller triggers signaling procedure to establish a D2D connection between devices $u$ and $v$. This algorithm is performed each time that a new device is attached to the network or every time that new SINR reports are received by the D2D controller.

\begin{algorithm}[!ht]
\footnotesize
\SetAlgoLined
\SetKwInOut{Input}{Input}\SetKwInOut{Output}{Output}
\Input{ $\gamma_{k}^{(v)}(u),\ \theta_{k}^{\prime}(u) \ \forall\ u,v \in \mathcal{D}$}
\Output{ D2D connection setup}
 \BlankLine
 \While{new SINR measurements from device $u$}{
 \For{$v \in \mathcal{D}$}{
  $\gamma_{\text{D2D}} = \min\left(\gamma_{k}^{(v)}(u),\gamma_{k}^{(u)}(v) \right) $;
  $\theta_{\text{BS}} = \max\left(\theta_{k}^{\prime}(u),\theta_{k}^{\prime}(v) \right) $;
   \BlankLine
  \eIf{$(\gamma_{\text{D2D}} \geq \delta_{\text{D2D}})$}{
  \BlankLine
   \tcp{D2D communication is possible \\ select the device relay}\label{cmt1}
    \BlankLine
        \eIf{$\theta_{\text{BS}}==\theta_{k}^{\prime}(u)$}
        {\BlankLine
        D$_{\text{relay}}$ = device $u$;}
        {D$_{\text{relay}}$ = device $v$;}
    \BlankLine    
    \tcp{Establish D2D connection}\label{cmt2}
    {$\bullet$ Triggers signaling}\;
   }{
    \tcp{D2D communication is not possible}\label{cmt3}
     {break}\;
  }
 }
 }
 \caption{D2D peering algorithm}
 \label{Algo:d2d1}
\end{algorithm}
\vspace{-1.0em}
\subsection{D2D Communication Phase}

Once a D2D connection is established between two devices, the D2D communication phase can be performed using dedicated radio resources (i.e. frequency band - Fig.~\ref{fig:int}). Therefore, in order to evaluate the global performances of D2D communication phase, we calculate the SINR as described below. \\\indent Let  $l$  be a D2D connection established between devices $u$ and $v$ over the channel $k$, we define $P_{k}(l)$ as the $\min\left(P^{(v)}_{k}(u),P^{(u)}_{k}(v)\right)$ and  $g_{k}(l)$ as its related channel gain. Finally, we assume that device $u$ is selected as relay node. \\\indent Therefore, we can compute the SINR  $\zeta_{k}(l)$ of the D2D connection $l$ as follows:
\begin{equation} \label{sinr3}
\zeta_{k}(l)=\frac{P_{k}(l) g_{k}(l)}{ \displaystyle\sum_{d \in \mathcal{L}, d \neq u,v} P^{(d)}_{k} (u) g^{(d)}_{k}(u)  + N_{th} }
\end{equation}
Let $W_{k}$ be the bandwidth in Hz of channel $k$. Thus, the achievable data rate for a D2D link $l$ which utilizes a channel $k$, can be calculated according to the well-known Shannon capacity formula:
\begin{equation} \label{eq:shannon}
\Upsilon_{k}(l)=W_{k}\log_{2} \left(1+\zeta_{k}(l)\right)
\end{equation}

\section{Wireless System Model}



\subsection{BS Antenna Pattern} \label{sub:bsantennapattern}

\begin{figure}[!ht]
\begin{subfigure}{.23\textwidth}
  \centering
  \includegraphics[width=.92\linewidth]{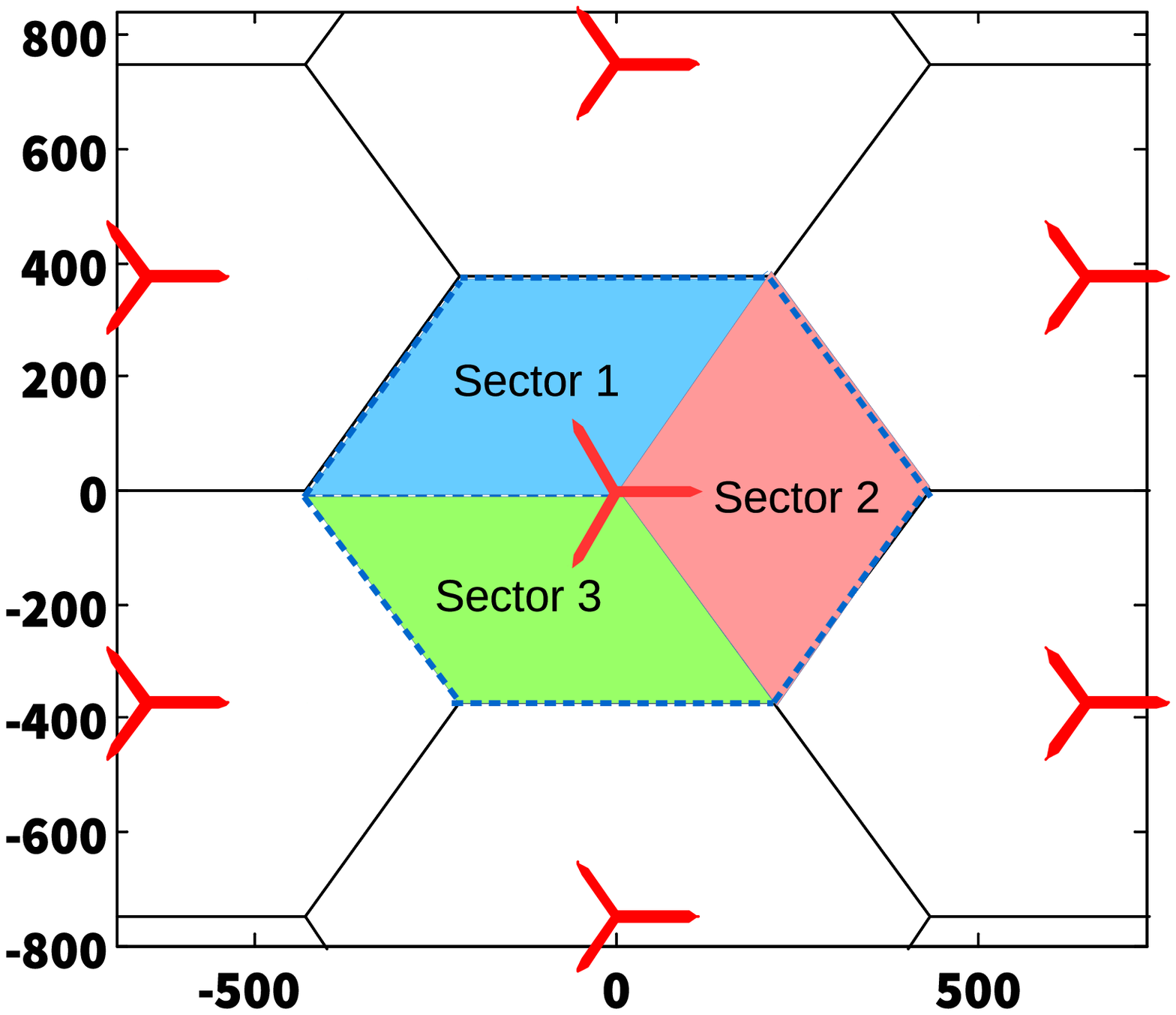}
  \caption{}
  \label{fig:cell}
\end{subfigure}%
\begin{subfigure}{.28\textwidth}
  \centering
  \includegraphics[width=.92\linewidth]{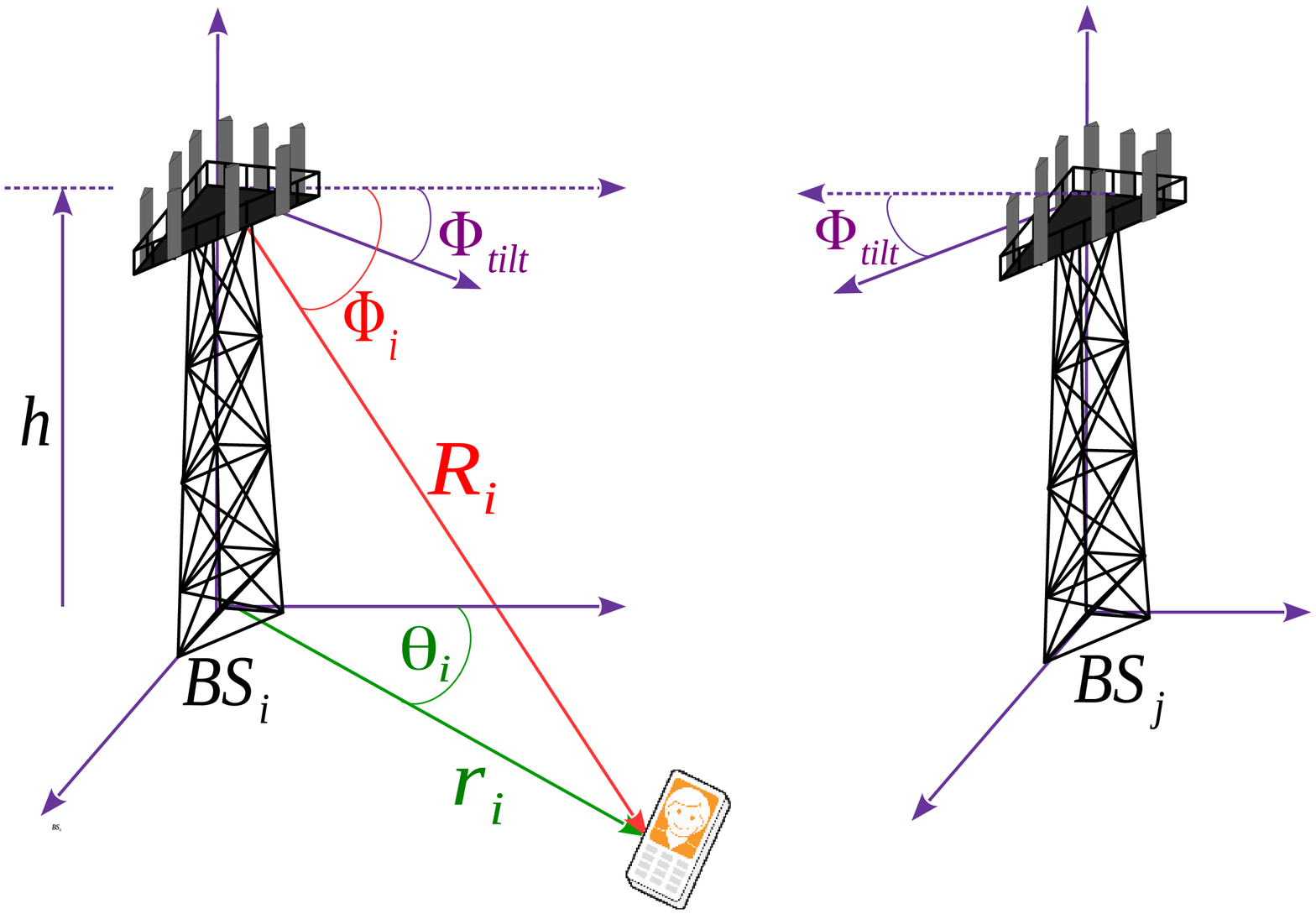}
  \caption{}
  \label{antennes3D-2}
\end{subfigure}
\vspace{-1.0em}
\caption{\footnotesize \textbf{(a)} Hexagonal network: location of the 3 sectors base stations in the plan. \textbf{(b)}  User equipment located at $(r_i, \theta_i, \Phi_i)$. It receives a useful power from antenna $i$ and interference power from antenna $j$.}
\vspace{-1.0em}
\label{fig:fig}
\end{figure}

Each site is constituted by 3 base stations, each one covering a sector, as shown in Fig.~\ref{fig:cell}.

According to the model of \cite{ITUR2009}, the antenna pattern applied to our scheme is computed as:
\begin{eqnarray} \label{eq:hv_pattern}
		A_{dB}(\theta, \phi) = -\min \left [ -(A_{h_{dB}}(\theta)+A_{v_{dB}}(\phi)), A_m \right ],
\end{eqnarray}
where $A_h(\theta)$ and $A_v(\phi)$ correspond respectively to the horizontal and the vertical antenna patterns.
The horizontal antenna pattern used for each base station (see Fig.~\ref{antennes3D-2}) is given by:
\begin{eqnarray} \label{eq:h_pattern}
		A_{h_{dB}}(\theta)= -\min \left[ 12 \left( \frac{\theta}{\theta_{3dB}} \right)^{2}, A_m \right ],
\end{eqnarray}
\noindent where:
\begin{itemize}
	\item $\theta_{3dB}$ is the half-power beamwidth (3 dB beamwidth);
	\item $A_m$ is the maximum attenuation.
\end{itemize}
The vertical antenna direction is given by:
\begin{eqnarray} \label{eq:v_pattern}
		A_{v_{dB}}(\phi)= -\min \left [ 12 \left( \frac{\phi-\phi_{tilt}}{\phi_{3dB}} \right)^{2}, A_m \right ],
\end{eqnarray}
\noindent where:
\begin{itemize}
	\item $\phi_{tilt}$ is the downtilt angle (see Fig.~\ref{antennes3D-2});
	\item $\phi_{3dB}$ is the 3 dB beamwidth.
\end{itemize}




\subsection{Energy Consumption Evaluation}

In order to evaluate the energy consumption we use the following approach.\\\indent In the regular scenario (non D2D communication), let $N$ be the total number of devices in the studied site. Therefore, we can compute the mean energy consumption $E^{\prime}$ (in J) over a TTI $T_{tti}$ as follows:
\begin{eqnarray} \label{eq:energy2}
	E^{\prime}=  P_{bs} N T_{tti}
\end{eqnarray}
\indent In the D2D  scenario, let $N_\text{d}$, $N_\text{d2d}$ and $N_\text{r}$ the number of regular users, D2D pairs and relay D2D users respectively. Moreover, let $P_{bs}$ and $P_{d2d}$ be the maximum BS transmitted power and the maximum  power transmitted by a device in a D2D link respectively. Therefore, we can compute the mean energy consumption $E$ (in J) over a Transmission Time Interval (TTI) $T_{tti}$ as follows:
\begin{eqnarray} \label{eq:energy}
	E =   \overbrace{P_{bs}(N_\text{d} + N_\text{r})T_{tti}}^\text{$E_{\text{1}}$} +  \overbrace{P_{d2d}N_\text{d2d} T_{tti}}^\text{$E_{\text{2}}$}
\end{eqnarray}
The mean energy consumption of the system is composed by \textbf{($E_{\text{1}}$)} energy consumption of communications between BS and device and \textbf{($E_{\text{2}}$)} energy consumption of D2D communications. It should be noted that communications between BS and a device are composed by regular communications and communications of D2D devices which act as relay nodes. 





\section{Numerical Results and Analysis}


We present hereafter the results obtained from numerical evaluation of two scenarios (with and without D2D). Our aim is twofold. We first focus on the discovery phase. We evaluate the impact of this phase on the pairing process, in particular in terms of D2D links number and reachable QoS. One of the main parameters is the required SINR threshold $\delta_{\text{D2D}}$, which determines the minimum QoS. Afterwards, we determine the impact of D2D communications, in terms of QoS and energy-efficiency, when a dedicated frequency bandwidth is used in downlink (DL). \\\indent In our analysis devices are randomly distributed in a cell of a 2D hexagonal based network (Fig.~\ref{fig:cell}). This hexagonal network is equipped by antennas which have a given height (h=30m in our analysis), in the third dimension.  All hexagon cells are identical in terms of devices placement. Therefore the analysis can be focused on the central cell for numerical calculation. Table~\ref{tab:simu1} summarizes simulation parameters.

\begin{table}[!ht]
\begin{center}
\begin{tabular}{|l|c|}
\hline 
Number of terminals & 400 by site (random positions) \\ 
\hline
eNB bandwidth &  from 20 to 1 MHz \\ 
\hline 
D2D bandwidth &  from 0 to 19 MHz \\ 
\hline 
Cell coverage radius & 500 m \\ 
\hline 
BS to device channel model & Cost231 \\ 
\hline 
D2D channel model & WINNER II~\cite{heino2010deliverable}\\ 
\hline 
BS Tx Power / Noise Figure & 46dBm / 5 dB \\ 
\hline 
D2D Tx Power / Noise Figure & From 20 to -20 dBm / 5 dB \\  
\hline 
D2D SINR threshold  & $\delta_{\text{D2D}}$ = 0 and 3 dB \\  
\hline 
DL carrier frequency & 2120 MHz \\  
\hline 
TTI & 1 ms \\  
\hline 
Antenna Gain  &   $G_0$ = 18 dB \\
\hline 
\end{tabular} 
\end{center}
\vspace{-0.5em}
\caption{Simulation parameters}
\vspace{-1.2em}
    \label{tab:simu1}   
\end{table}

 We present two types of comparisons. We establish the Cumulative Distribution Functions (CDF) of the throughput based on (\ref{eq:shannon}) for D2D communications and for devices directly connected to their serving BS. The CDF of the throughput characterizes the outage probability and the performance distribution. Indeed, these curves provide information related to network characteristics such as QoS and coverage, outage probability and performance that can be reached by users in the system. A second analysis is focused on the analysis of the energy used, and its impact of number of established D2D pairs, mean experimented throughput and its efficiency.

\begin{figure*}[!t]
\includegraphics[width=\textwidth]{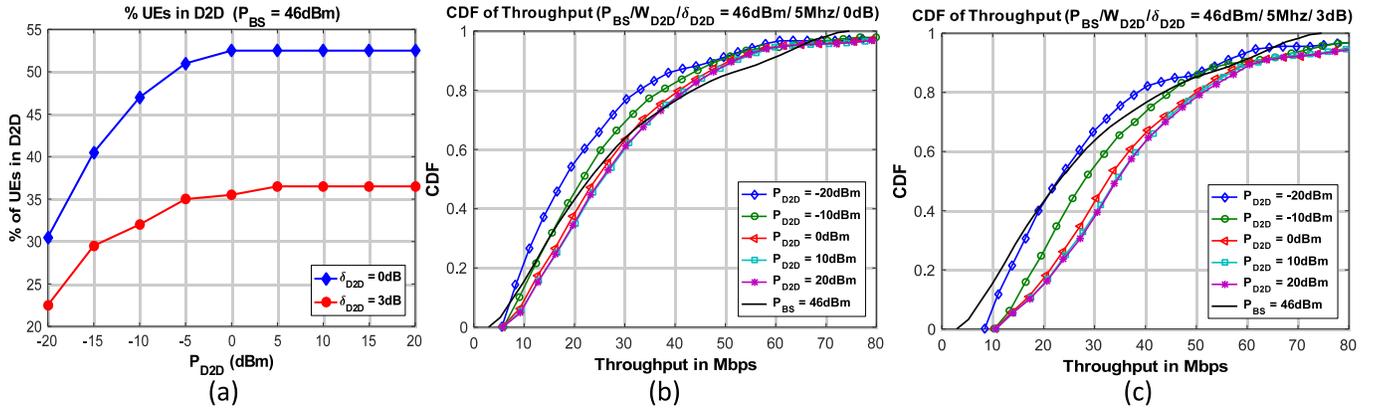}
\vspace{-1.4em}
\caption{\footnotesize (a)  Percentage of D2D users according to different D2D transmitting powers  (b) CDF of Throughput for D2D connections with $\delta_{D2D}$= 0~dBm  (c) CDF of Throughput for D2D connections with $\delta_{D2D}$= 3~dBm}
\label{fig:d2d-1}
\vspace{-0.8em}
\end{figure*}
\begin{figure*}[!ht]
\includegraphics[width=\textwidth]{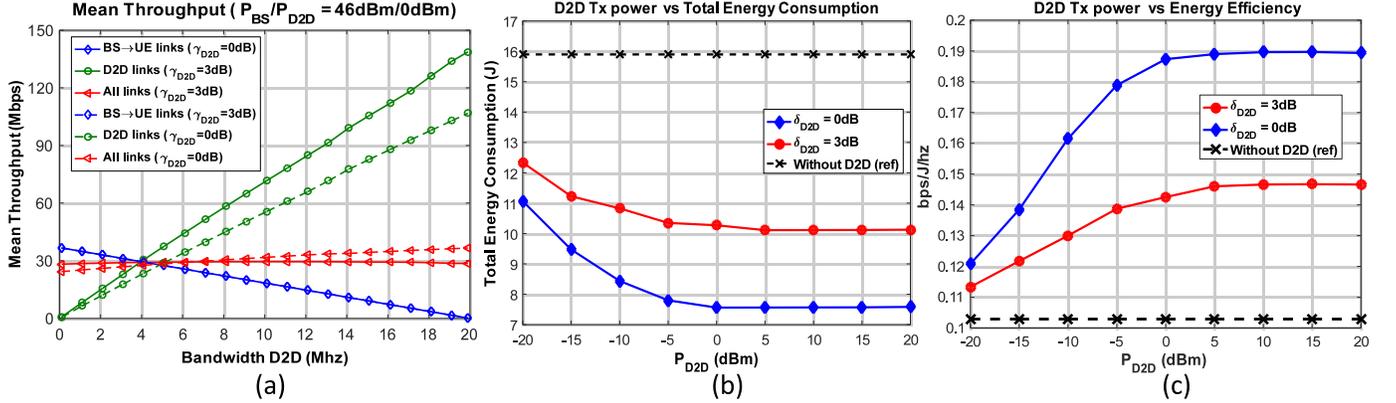}
\vspace{-1.5em}
\caption{\footnotesize (a) Mean Throughput of D2D scenario  according to D2D bandwidth for $\delta_{D2D}$= 0~dB and 3~dB (b) Total energy consumption according to different D2D transmitting powers with $P_{BS}$= 46~dBm  (c) Energy-Efficiency according to different D2D transmitting powers for $\delta_{D2D}$= 0~dB and 3~dB, $P_{BS}$= 46~dBm and $W_{D2D}$= 5~Mhz. Black dotted curves in (b) and (c) represent scenarios without D2D communications.}
\label{fig:d2d-2}
\vspace{-1.2em}
\end{figure*}

\subsection{Discovery Phase Analysis}

The curves drawn in Fig.~\ref{fig:d2d-1}(a) present the percentage of D2D connections performed during the discovery phase according to different D2D transmitting powers, in the cases where the minimum SINR requested for a D2D connection are $\delta_{D2D}=0$~dB and 3~dB. It's worth recalling that the threshold $\delta_{\text{D2D}}$ allows to guarantee a minimum throughput level and it is used on the discovery phase. It first can be observed that the percentage of D2D connections increases when the D2D transmitting power increases. Moreover, reducing D2D threshold ($\delta_{D2D}$) from 3~dB to 0~dB, increases by 50\% the number of devices involved in a D2D communication, when the D2D transmitting power is higher than 0~dBm. Therefore, the number of D2D connections for $\delta_{D2D}=0$~dB is significantly greater than for $\delta_{D2D}=3$~dB. It can also be observed that allowing a D2D transmitting power higher than 0 dBm has no impact on the variation of the number of D2D connections (around  53\% when $\delta_{D2D}=0$~dB and around 36\% when $\delta_{D2D}=3$~dB). This is a consequence of the interference induced between devices since they use the same frequency bandwidth. \\\indent The following is the analysis of communication phase, where the impact of the discovery phase is evidenced.

\subsection{Communication Phase Analysis}

\subsubsection{Performance, outage probability and coverage}

The curves presented in Fig.~\ref{fig:d2d-1}(b) and Fig.~\ref{fig:d2d-1}(c) compare the CDF of throughput for devices directly connected to their serving BS (black curves) to the CDF representing D2D connections (colored curves). In Fig.~\ref{fig:d2d-1}(b) the minimum SINR requested for a D2D connection is chosen to $\delta_{D2D}=0$~dB and in Fig.~\ref{fig:d2d-1}(c) $\delta_{D2D}=3$~dB. \\\indent  It can be observed that in Fig.~\ref{fig:d2d-1}(b), regular connections outperform D2D connections when the D2D transmitted power is less or equal to -10~dBm. Moreover, when the D2D transmitted power is greater or equal to 0~dBm (i.e. 0~dBm, 10~dBm and 20~dBm) the CDF of D2D connections are close one to another. Therefore, it is sufficient to allocate a low D2D transmission power (i.e. 0~dBm) to reach a higher level of performance. \\\indent In case of Fig.~\ref{fig:d2d-1}(c), regular connections outperform D2D connections only for -20~dBm  of D2D transmitted power. The performance distribution shows that about 15\% of users have a throughput lower than 10~Mbps (black curve) and 100\% of D2D users reach a throughput higher than 10~Mbps (colored curves) in the case of D2D connections. \\\indent  In both cases ($\delta_{D2D}=0$~dB and 3~dB) the D2D transmitted power of 0~dBm shows better performances than regular connections. This means that using 0~dBm of transmitted power allows D2D connections to reach a higher level of performance, coverage and QoS than for devices directly connected to the BS. To establish these curves, the transmitted power allocated to BS is 46~dBm.  For example, it can be observed in Fig.~\ref{fig:d2d-1}(b) that for an outage probability of 10\% (value 0.1 of the CDF), the throughput reached without D2D is 7~Mbps. With D2D connections, the throughput reaches about 18~Mbps, which represents an increase of 11 Mbits/s i.e. about 150\% in terms of throughput. \\\indent The curves presented in Fig.~\ref{fig:d2d-1}(c) compare the average throughput of D2D connections, regular connections and the global system when different bandwidth sizes of total available bandwidth (i.e. 20 Mhz) are assigned to the overlay D2D communication. \\\indent In case of D2D bandwidth greater or equal to 5~Mhz, it can be observed that the global system performance in terms of throughput is higher when $\delta_{D2D}=0$~dB. Otherwise, the case of $\delta_{D2D}=3$~dB has better global performances.

\subsection{Discovery Phase Impact on Energy-Efficiency}

The curves drawn in Fig.~\ref{fig:d2d-2}(b) present the mean energy consumption of the overall system based on Equations (\ref{eq:energy2}) (without D2D) and (\ref{eq:energy}) (with D2D) according to different D2D transmitting powers, in the cases where $\delta_{D2D}=0$~dB and 3~dB (\emph{discovery phase}). It first can be observed that the mean energy consumption decreases when the D2D transmitting power increases. This is due to the fact that the percentage of D2D connections increases when the D2D transmitting power increases, and that D2D connections consume lower energy than regular communications. It can also be observed that D2D transmitted powers greater than 0~dBm have no impact on the variation of the energy consumption. Indeed, it is directly related to the percentage of D2D connections, which does not vary, too. It is important to highlight that D2D scenarios outperform reference scenario (without D2D communications - black curve). Moreover, in the best case ($P_{D2D}=0$ dBm) the energy consumption is reduced by over 50~\% ($\delta_{D2D}=0$~dB - blue curve). \\\indent The curves drawn in Fig.~\ref{fig:d2d-2}(c) present the energy-efficiency in terms of bps/J/Hz according to different D2D transmitting powers, in the cases where $\delta_{D2D}=0$~dB and 3~dB. It first can be observed that the energy-efficiency increases when the D2D transmitting power increases. Moreover, the increase is higher for $\delta_{D2D}=0$~dB than for $\delta_{D2D}=3$~dB. It can also be observed that D2D transmitted powers greater than 0~dBm have a negligible impact on the variation of the energy-efficiency. It is important to highlight that D2D scenarios outperform reference scenario (without D2D communications - black curve), which is around 0.103~ bps/J/Hz. Moreover, in the best case ($P_{D2D}=0$ dBm) the energy efficiency is improved by over 80~\% ($\delta_{D2D}=0$~dB - blue curve). \\\indent Therefore, in the light of the above analysis, it can be inferred that some configurations of D2D systems can guarantee a good QoS level while improving energy-efficiency. In our numerical evaluation, the best configuration corresponds to D2D transmitting power $P_{D2D}=0$~dBm, a minimum SINR requested for a D2D connection $\delta_{D2D}=$0~dB,  and a frequency bandwidth of 5~Mhz for D2D communications. \\\indent In summary, the numerical evaluation shows improvements provided by D2D communications in terms of QoS, when appropriated parameters are used. As is the case of the SINR threshold $\delta_{\text{D2D}}$, which is a key parameter on D2D communications and used during the discovery phase. This QoS improvement allows to increase the achievable throughput and the coverage extension as well as to reduce the outage probability and the energy consumption. 




\section{Conclusion and Future Work}

In this paper, a D2D centralized architecture is considered, in line with current trends for 5G mobile networks definition. We have proposed a performance evaluation of overlaying D2D communication in terms of QoS and energy-efficiency, by taking into account both the discovery and the communication phases. An opportunistic pairing algorithm is developed, which aims at improving perceived SINR. The relevant impact of the discovery phase has been analyzed, in terms of energy-efficiency as well as QoS. Numerical evaluation reveals interesting performance improvements thanks to D2D communication, since it can guarantee a good QoS level while enhancing energy-efficiency. These performance improvements are directly related to the choice of the D2D system configuration (i.e. $P_{D2D}$, $\delta_{D2D}$). \\\indent In a future work dynamic traffic analysis will also be considered as well as device mobility scenarios. 

\bibliographystyle{IEEEtran}
\bibliography{IEEEabrv,biblio}

\end{document}